# Theory of aces: high score by skill or luck?

M.V. Simkin and V.P. Roychowdhury
*Department of Electrical Engineering, University of California, Los Angeles, CA 90095-1594*

We studied the distribution of WWI fighter pilots by the number of victories they were credited with along with casualty reports. Using the maximum entropy method we obtained the underlying distribution of pilots by their skill. We find that the variance of this skill distribution is not very large, and that the top aces achieved their victory scores mostly by luck. For example, the ace of aces, Manfred von Richthofen, most likely had a skill in the top quarter of the active WWI German fighter pilots, and was no more special than that. When combined with our recent study [10], showing that fame grows exponentially with victory scores, these results (derived from real data) show that both outstanding achievement records and resulting fame are mostly due to chance.

During the "Manhattan project" (the making of nuclear bomb), physicist Enrico Fermi asked General Leslie Groves, the head of the project, what is the definition of a "great" general [1]. Groves replied that any general who had won five battles in a row might safely be called great. Fermi then asked how many generals are great. Groves said about three out of every hundred. Fermi conjectured that if the chance of winning one battle is 1/2 then the chance of winning five battles in a row is $1/2^5 = 1/32$. "So you are right, General, about three out of every hundred. Mathematical probability, not genius."

Similarly to a great general, an ace is a fighter pilot who achieved five or more victories. Can the latter be explained by simple probability, like the former? At first glance this does not appear to be so, as some aces scored way too many victories. For example, the probability to achieve by pure chance Manfred von Richthofen's 80 victories is $1/2^{80} \approx 10^{-24}$. One is tempted to conclude that high-scoring aces had outstanding skills.

A more careful analysis proves this conclusion wrong. During WWI British Empire Air Forces fully credited their pilots for *moral*[1] victories (Ref. [2], p.6). It is not that unlikely to achieve five moral victories if you can have five moral defeats in between. In addition British Air Force fully credited their pilots for shared victories (Ref. [2], p.8). That is if e.g. three British airplanes shot one German airplane – all three were credited with a victory. The French did not count moral victories, but allowed for shared ones (Ref. [3], p.6). The Americans were either under French or British command and had the corresponding rules applied to them. In contrast, the Germans had ideal scoring system (Ref. [4], p.6-7). They did not count moral victories. The opponent aircraft had to be either destroyed or forced to lend on German territory and its crew taken prisoners. They did not allow shared victories as well[2]. This was in theory. In practice, however, military historians "have found a number of 'victories' where, say, three Allied aircraft have been claimed and credited when there is absolutely no doubt that only one or two of those Allied planes were lost" (Ref. [4], p.7). This means that in reality some moral or shared victories were counted by the Germans.

---

[1] For example, forcing the enemy aircraft to land within enemy lines, driving it down "out of control", or driving it down in damaged condition (Ref. [2], p.6).

[2] This brought another problem. It happened that there were two or more claims for one destroyed opponent aircraft. Military historians had found that "In some of these cases rank or being a higher scoring ace helped win the decision over a more lowly pilot" (Ref. [4], p.7). Several such cases are documented in Ref. [5]: Vizefeldwebel (Sergeant-Major) Boldt claimed a victory, but it was also claimed by and awarded to Ltn von Schönebeck (Ref. [5], p.108); Vizefeldwebel (Sergeant-Major) Hegeler claimed a victory, but it was also claimed by and awarded to Ltn d R Müller (Ref. [5], p.157). This phenomenon, if widespread, can alone generate aces through the cumulative advantage

Ref. [5] contains the list of all German WWI fighter pilots, with all of their victories and casualties. The total number of credited victories is 6759[3]. The number of casualties, however, is a lot smaller[4]. They amount to 618 KIA (killed in action), 52 WIA/DOW (wounded in action and later died of wounds), 140 POW (prisoner of war), and 431 WIA (wounded in action and survived). According to the official German scoring system, for a pilot to be credited with a victory his opponent should be killed or taken prisoner. Let us compute the number of defeats suffered by the Germans using their own scoring system for victories. Obviously, KIA, WIA/DOW, and POW should be counted as defeats. These add up to 810. This is by a factor of 8.3 less than the number of credited victories. We are not supposed to include WIA in defeats if we wish to follow the German scoring system. However, even if we count all of the WIA as defeats we get 1,241 defeats, which is still by a factor of 5.4 less than the number of credited victories.

We don't know for sure why the number of victories exceeds the number of casualties by such a large factor, but can suggest several possible reasons:

- Moral and shared victories.
- Aces flew fighter-planes, while their opponents often were less well armed aircraft.
- German Air Force fought mostly defensive war behind their front lines [4]. So, if a German aircraft was shot down, it could land on their territory. In contrast, when Allied aircraft was shot down, it had to land on the enemy territory and its pilot was taken prisoner.
- The Germans were better.

Fortunately, we don't need to know the exact reason to compare German fighter pilots between themselves. Let us, given the statistics of defeats and victories, compute the probability to get Richthofen's score. German pilots were credited with 6759 victories (this number probably includes moral and shared victories). Germans also recorded 810 defeats. The total number of engagements was probably not 6759 + 810, but 6759 + 810 + X. Here X is the unknown number of moral defeats. As long as moral defeat does not affect the ability of a pilot to participate in further battles we don't need to know X. We will call a "recorded engagement" an engagement which resulted in either credited victory or in a defeat. The rate of defeat in recorded engagements is $r = \frac{810}{6759 + 810} \approx 0.107$. The probability of 80 victories in a row is $(1-r)^{80} \approx 1.17 \times 10^{-4}$. The probability that at least one of 2894 German fighter pilots will achieve 80 or more victories is $1 - (1 - 1.17 \times 10^{-4})^{2894} \approx 0.29$. Richthofen's score is thus within the reach of chance. We can also compute the probability distribution of the victory scores, assuming that everyone fights until he gets killed. The probability to win $n$ fights and lose the next is:

---

mechanism. However, we have no evidence that this practice was widespread, and will ignore its effect in this study.

[3] This number is the sum of 5050 victories credited to aces and 1709 victories credited to non-ace pilots. The first number is accurate (in the sense that no new error was introduced in this study), as aces victory scores are available in electronic format (for example on this website: http://www.theaerodrome.com/aces/). The second number is a result of the hand-count using the listing in Ref. [5], so some error was most likely introduced.

[4] The casualties, which are listed on pages 345-357 of Ref. [5], were manually counted.

$$P(n) = (1-r)^n r \tag{1}$$

Figure 1 shows the result of Eq.(1) (with $r = 0.107$) compared with the actual distribution of the victory scores (which are given in Table 1). While the agreement is not perfect, it is clear that chance can account for most of the variance in the numbers of victories.

Apart from not leading to a quantitative agreement with the data, the above simple analysis assumes that fighter pilots always fight until they get killed. In reality many of them did not even look for a fight. There were over eight hundred German fighter-pilots who did not score a single victory and also were never wounded or shot down. Also many pilots with just few victories survived the war. In many of such cases they joined the Air Force shortly before the end of the war.

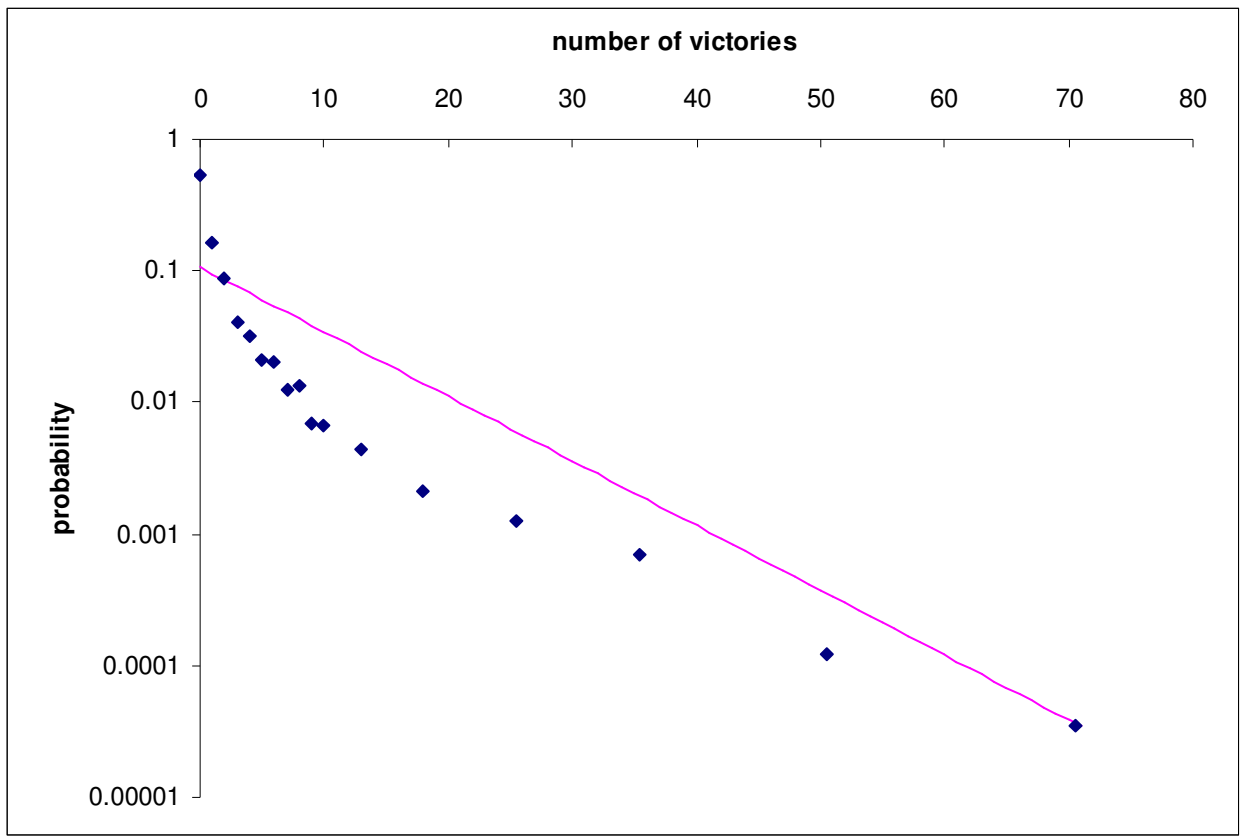

**Figure 1.** The distribution of German WWI fighter-pilots by the number of victories, computed using the data in Table 1, is shown by rhombs. The line is the predicted distribution, computed using Eq.(1) with $r = 0.107$.

**Table 1.** Distribution of pilots by number of victories.

| number of credited victories | total number of pilots | un-defeated | defeated in the next fight | number of credited victories | total number of pilots | un-defeated | defeated in the next fight |
|---|---|---|---|---|---|---|---|
| 0 | 1567 | 1133 | 434 | continued | | | |
| 1 | 469 | 331 | 138 | 23 | 1 | 1 | 0 |
| 2 | 252 | 182 | 70 | 24 | 2 | 1 | 1 |
| 3 | 120 | 90 | 30 | 25 | 3 | 1 | 2 |
| 4 | 94 | 64 | 30 | 26 | 3 | 2 | 1 |
| 5 | 60 | 47 | 13 | 27 | 6 | 4 | 2 |
| 6 | 59 | 47 | 12 | 28 | 5 | 4 | 1 |
| 7 | 36 | 27 | 9 | 29 | 1 | 1 | 0 |
| 8 | 39 | 27 | 12 | 30 | 5 | 3 | 2 |
| 9 | 20 | 15 | 5 | 31 | 2 | 1 | 1 |
| 10 | 19 | 13 | 6 | 32 | 2 | 2 | 0 |
| 11 | 15 | 11 | 4 | 33 | 3 | 2 | 1 |
| 12 | 18 | 12 | 6 | 35 | 5 | 5 | 0 |
| 13 | 11 | 5 | 6 | 36 | 3 | 2 | 1 |
| 14 | 4 | 3 | 1 | 39 | 2 | 1 | 1 |
| 15 | 15 | 9 | 6 | 40 | 3 | 2 | 1 |
| 16 | 6 | 4 | 2 | 43 | 1 | 1 | 0 |
| 17 | 8 | 7 | 1 | 44 | 2 | 2 | 0 |
| 18 | 3 | 2 | 1 | 45 | 1 | 0 | 1 |
| 19 | 4 | 3 | 1 | 48 | 2 | 1 | 1 |
| 20 | 10 | 6 | 4 | 54 | 1 | 0 | 1 |
| 21 | 4 | 1 | 3 | 62 | 1 | 1 | 0 |
| 22 | 6 | 3 | 3 | 80 | 1 | 0 | 1 |

A better way to address the problem is to study the probability of defeat as a function of the number of previous victories. Table 2 shows the statistics of casualties (KIA + WIA/DOW +POW) as a function of the number of previous victories. For example, 434 pilots were defeated before they achieved a single victory. At the same time 1321 pilots achieved one or more victory. This makes the rate of defeat in the first fight $\frac{434}{434+1327} \approx 0.247$. Similarly 30 people were defeated after they achieved 4 victories, while 392 pilots achieved 5 or more victories (and became aces). This makes the rate of defeat in the fifth fight $\frac{30}{30+392} \approx 0.071$.

The rate of defeat computed this way using the data of Table 2 is shown in Fig.2. The rate drops strongly for the first few fights, but appears not to change after about ten fights. The reduction in rate of defeat can be explained assuming that pilots have different innate rate of defeat, which depends on their skill. The unskilled ones get killed at a higher rate, and, as we progress to higher number of fights, the average skill increases. Variance in innate skill is one possible explanation. Another explanation is that the pilots simply get more experienced. A hard-core idealist can attribute all the decrease of the defeat rate with the increase of the fight number to learning. In reality both factors play role and their relative contributions are impossible to determine. In the following analysis we assume that a defeat rate is innate to a pilot and does not change with the number of fights he participates in.

**Table 2.** Numbers of defeated and winning pilots as functions of the fight number.

| fight number | number of defeats | number of victories | fight number | number of defeats | number of victories |
|---|---|---|---|---|---|
| 1 | 434 | 1327 | | *continued* | |
| 2 | 138 | 858 | 42 | 0 | 9 |
| 3 | 70 | 606 | 43 | 0 | 9 |
| 4 | 30 | 486 | 44 | 0 | 8 |
| 5 | 30 | 392 | 45 | 0 | 6 |
| 6 | 13 | 332 | 46 | 1 | 5 |
| 7 | 12 | 273 | 47 | 0 | 5 |
| 8 | 9 | 237 | 48 | 0 | 5 |
| 9 | 12 | 198 | 49 | 1 | 3 |
| 10 | 5 | 178 | 50 | 0 | 3 |
| 11 | 6 | 159 | 51 | 0 | 3 |
| 12 | 4 | 144 | 52 | 0 | 3 |
| 13 | 6 | 126 | 53 | 0 | 3 |
| 14 | 6 | 115 | 54 | 0 | 3 |
| 15 | 1 | 111 | 55 | 1 | 2 |
| 16 | 6 | 96 | 56 | 0 | 2 |
| 17 | 2 | 90 | 57 | 0 | 2 |
| 18 | 1 | 82 | 58 | 0 | 2 |
| 19 | 1 | 79 | 59 | 0 | 2 |
| 20 | 1 | 75 | 60 | 0 | 2 |
| 21 | 4 | 65 | 61 | 0 | 2 |
| 22 | 3 | 61 | 62 | 0 | 2 |
| 23 | 3 | 55 | 63 | 0 | 1 |
| 24 | 0 | 54 | 64 | 0 | 1 |
| 25 | 1 | 52 | 65 | 0 | 1 |
| 26 | 2 | 49 | 66 | 0 | 1 |
| 27 | 1 | 46 | 67 | 0 | 1 |
| 28 | 2 | 40 | 68 | 0 | 1 |
| 29 | 1 | 35 | 69 | 0 | 1 |
| 30 | 0 | 34 | 70 | 0 | 1 |
| 31 | 2 | 29 | 71 | 0 | 1 |
| 32 | 1 | 27 | 72 | 0 | 1 |
| 33 | 0 | 25 | 73 | 0 | 1 |
| 34 | 1 | 22 | 74 | 0 | 1 |
| 35 | 0 | 22 | 75 | 0 | 1 |
| 36 | 0 | 17 | 76 | 0 | 1 |
| 37 | 1 | 14 | 77 | 0 | 1 |
| 38 | 0 | 14 | 78 | 0 | 1 |
| 39 | 0 | 14 | 79 | 0 | 1 |
| 40 | 1 | 12 | 80 | 0 | 1 |
| 41 | 1 | 9 | 81 | 1 | 0 |

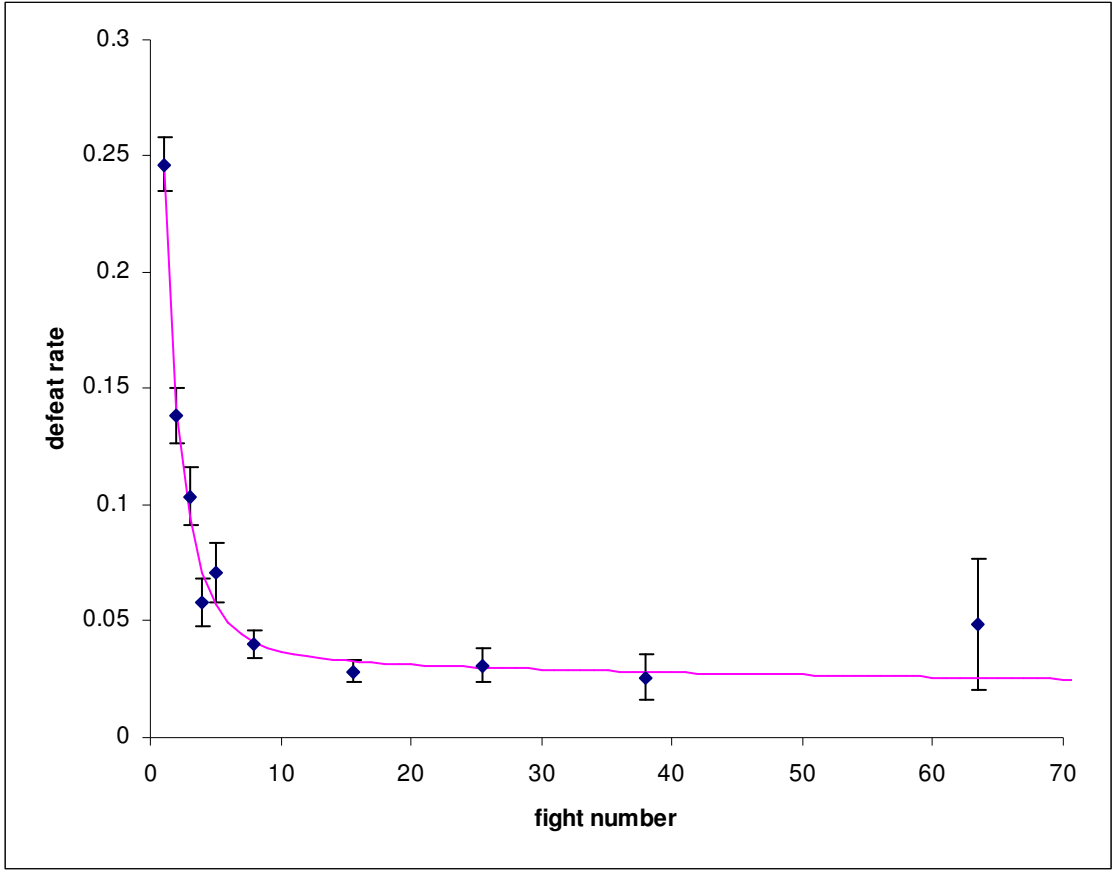

**Figure 2.** The actual defeat rate as a function of fight number, determined using the data of Table 2, is shown by rhombs. The line represents the same rate, theoretically computed using the distribution of the rate of defeat in the pool of pilots given in Fig. 3. The error bars are of the size of one standard deviation, which corresponds to 68% confidence interval.

If the distribution of the innate defeat rates is $p(r)$ then the average rate of defeat in first fight is $\bar{r}_1 = \int_0^1 r\, p(r)\, dr$. The probability distribution of defeat rate of pilots surviving the first fight is $p_1(r) = \dfrac{(1-r)p(r)}{\int_0^1 (1-r')p(r')dr'}$. The rate of defeat in the second fight is $\bar{r}_2 = \int_0^1 r p_1(r)\, dr = \dfrac{\int_0^1 r(1-r)p(r)dr}{\int_0^1 (1-r)p(r)dr}$.

In general, the probability distribution of defeat rates of pilots, surviving $n$ fights, is

$$p_n(r) = \dfrac{(1-r)^n p(r)}{\int_0^1 (1-r')^n p(r')dr'}, \qquad (2)$$

and the rate of defeat in $n$th fight is:

$$\bar{r}_n = \frac{\int_0^1 r(1-r)^{n-1} p(r)dr}{\int_0^1 (1-r)^{n-1} p(r)dr} = 1 - \frac{\int_0^1 (1-r)^n p(r)dr}{\int_0^1 (1-r)^{n-1} p(r)dr}. \qquad (3)$$

Obviously, $\bar{r}_n$, given by Eq.(3), monotonically decreases with $n$. When the minimum defeat rate in the distribution $p(r)$ is greater than zero, $\bar{r}_n$ approaches this rate at some value of $n$ and then decreases no further. Fig. 2 suggests that this minimum defeat rate is around 3%.

One can use Eq.(3) to compute the defeat rates for trial distributions, $p(r)$, and search for the distribution which best fits the data of Fig. 2. However, a better way to find $p(r)$ is to use the method of maximum likelihood [6]. Let us consider an undefeated pilot with $n$ victories. We assume that he was selected at random from a pool where defeat rates are distributed according to $p(r)$. The probability that a randomly selected pilot, who participates in $n$ battles, wins all of them is:

$$P(n) = \int_0^1 dr p(r)(1-r)^n.$$

Similarly, the probability that a randomly selected pilot wins $n$ battles and loses the next is:

$$P^d(n) = \int_0^1 dr p(r)(1-r)^n r = P(n) - P(n+1).$$

Now the probability to get the whole set of data in Table 1 is given by the likelihood function:

$$f = \prod_n (P(n))^{N(n)} (P^d(n))^{N^d(n)}$$

where $N(n)$ and $N^d(n)$ are the numbers of undefeated and defeated pilots with $n$ victories. We should find the distribution $p(r)$ which maximize this function.

In computations it is more convenient to work with the logarithm of the likelihood function:

$$\ln(f) = \sum_n [N(n)\ln(P(n)) + N^d(n)\ln(P^d(n))]. \qquad (4)$$

The distribution, $p(r)$, obtained by maximizing $f$ (see the Appendix) is shown in Fig. 3 by rhombs. It looks irregular.

The maximum likelihood estimation we just performed assumed that all possible distributions, $p(r)$, are, a priori, equally probable. The Maximum Entropy Principle [7] provides a more reasonable way of assigning a priori probabilities to distributions. As we are inferring a *probability* distribution then the relevant entropy is the information (Shannon) entropy [8]:

$$s = -\int_0^1 dr p(r) \ln(p(r)) \qquad (5)$$

A priori probability of a given probability distribution, $p(r)$, is $\propto e^s$ [7]. The combined probability of realizing a particular distribution and that this distribution produces the observed data is $\sim e^s \times f$.

This is the quantity which should be maximized, or, alternatively, its logarithm, $\ln(f)+s$, which is more convenient. The result of this maximization is shown in Fig. 3 by a line. The defeat rate as a function of fight number, computed using this distribution and Eq.(3) is shown in Fig.2 by a line.

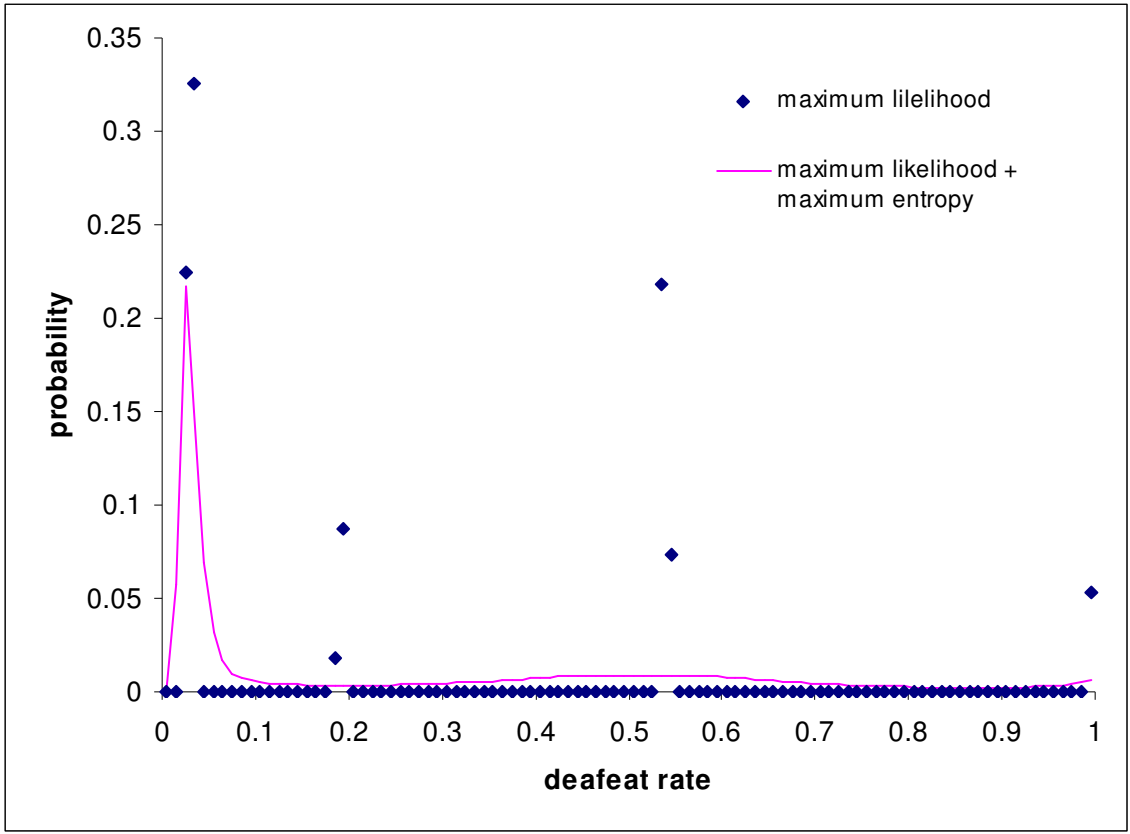

**Figure 3**. Distribution of the rate of defeat computed using maximum likelihood method (rhombs). Solid line is the distribution computed by combining the maximum likelihood and the maximum entropy method.

Now we can use $p(r)$ to do Bayesian inference [6] for intrinsic defeat rate of any given pilot (including those who were never defeated[5]). We will use $p(r)$ as a prior distribution of defeat rate and will make an estimate of pilot's defeat rate based on this prior distribution and actual number of fights he won and lost. For example, if we don't know how many fights a pilot had won, then all we can say is that the probability distribution of his defeat rate is $p(r)$. If we know that he won $n$ fights, then the probability distribution of his defeat rate is the probability distribution of defeat rates of pilots, who won $n$ fights, which is given by Eq.(2). The same inference for defeat rate for pilots, who won $n$ fights and were defeated in the next, can be obtained similarly:

$$p_n^d(r) = \frac{r(1-r)^n p(r)}{\int_0^1 r'(1-r')^n p(r')dr'} \qquad (6)$$

---

[5] Similar approach was previously used to estimate the true dropped calls rates when no dropped calls happened during the test [9].

The inference for five representative pilots, computed using Eqs. (2) and (6) is given in Fig. 4. In particular, Manfred von Richthofen most likely had the intrinsic defeat rate of 2.5%. According to the distribution of intrinsic defeat rates shown in Fig. 3 about 27% of pilots have the defeat rate of 2.5% or lower. This means that MvR is most likely merely in top 27% according to his skill.

Note that we completely neglected the effects of learning in our analysis. It is clear that at least part of the variation in perceived ability is due to the fact that pilots are getting more experienced as they participate in more and more fights. Thus taking into account the effects of learning will make the variance in innate ability only less. Consequently, our estimate of the uniqueness of MvR is an upper bound.

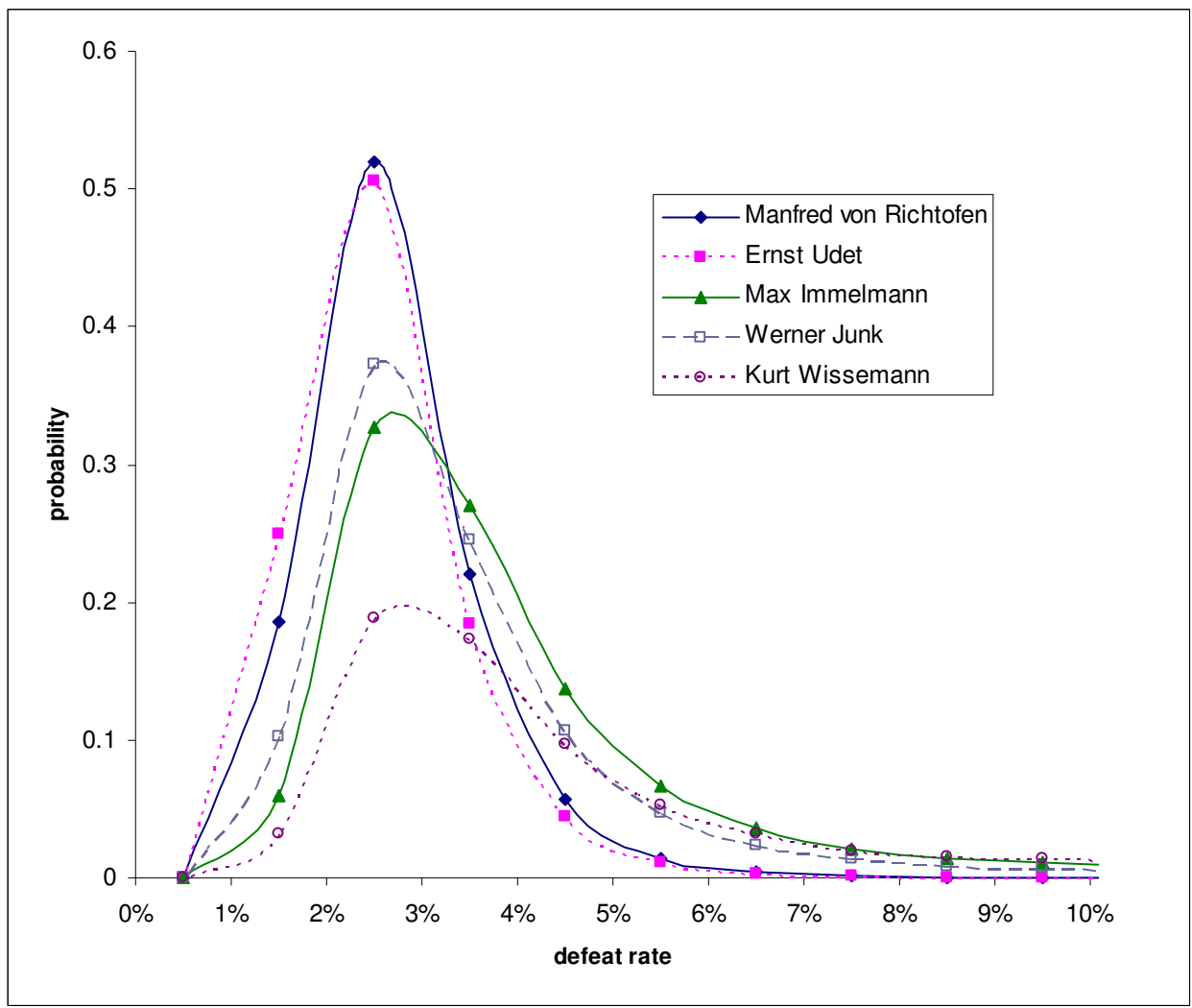

**Figure 4.** Bayesian inference for the defeat rate of five aces, computed using the distribution, shown in Fig. 3, and Eqs.(2) and (6). Undefeated aces (number of victories in brackets): Ernst Udet (62), Werner Junck (5). Defeated aces: Manfred von Richthofen (80), Max Immelmann (15), Kurt Wissemann (5).

In our previous paper on the theory of aces [10] we found a strong correlation between the logarithm of fame (measured in numbers of Google hits) and the achievement (number of victories), suggesting that fame grows exponentially with achievement. In other words fame gives increasing return on achievement, but still is determined by this achievement. This would be acceptable if achievement was proportional to skill. However, now we have shown that the difference in the number of victories is mostly due to chance. This means that the fame in the end is due to chance. There were a couple of papers ([11], [12]) which speculated, using arguments akin to the one by Fermi in the beginning of the article, that people can be perceived as having extraordinary ability when in reality they are simply lucky. However, this paper is the first one, which argues it using real data.

## Appendix (The maximization algorithm)

The method used is as follows. The defeat rates are discretized: $r_k = 0.005 + 0.01 \times k$, $k = 0,\ldots,99$. The probability also is discretized in 1000 units of 0.001. These 1000 units of probability are initially distributed at random over the 100 defeat rates. Then we use the following maximization algorithm. Start with $k = 0$. Reduce the number of probability units at $k = 0$ (if there is any) by one. Tentatively move this probability unit to $k = 0,\ldots,99$ and compute $\ln(f)$ in each case. Stick with the move which maximizes $\ln(f)$. Proceed to $k = 1$ and repeat the procedure and so on. After $k = 99$ go back to $k = 0$ and repeat the whole cycle. Stop when no further move increases $\ln(f)$.

The whole run was repeated 500 times each time starting with a different random probability distribution. These runs ended up in 33 distinct "local" maximums. The result of the most successful run is shown in Fig. 3. Note, that all of the runs produced very similar probability distributions. The best run resulted in the likelihood, exceeding that of the worst run by the factor of only 1.002.

The maximum entropy case was treated exactly the same way. The only difference was that $\ln(f)$ was replaced with $\ln(f) + \ln(s)$. One hundred runs ended up in the same maximum.